\documentclass[twocolumn]{jpsj2}

\title{%
Conductance Fluctuations in Disordered Wires
with Perfectly Conducting Channels
}

\author{%
Yositake {\sc Takane}$^{1}$ and Katsunori {\sc Wakabayashi}$^{1,2}$
}

\inst{%
$^{1}$Department of Quantum Matter, Graduate School of Advanced Sciences of Matter, Hiroshima University, Higashi-Hiroshima 739-8530, Japan \\
$^{2}$PRESTO, Japan Science and Technology Agency (JST), 4-1-8 Honcho,
Kawaguchi, Saitama 332-0012, Japan
}

\recdate{ \hspace{50mm} }

\abst{%
We study conductance fluctuations in disordered quantum wires
with unitary symmetry focusing on the case in which
the number of conducting channels in one propagating direction
is not equal to that in the opposite direction.
We consider disordered wires with $N+m$ left-moving channels
and $N$ right-moving channels.
In this case, $m$ left-moving channels become perfectly conducting,
and the dimensionless conductance $g$ for the left-moving channels
behaves as $g \to m$ in the long-wire limit.
We obtain the variance of $g$ in the diffusive regime by using
the Dorokhov-Mello-Pereyra-Kumar equation for transmission eigenvalues.
It is shown that the universality of conductance fluctuations breaks down
for $m \neq 0$ unless $N$ is very large.
}

\kword{%
perfectly conducting channel, universal conductance fluctuations,
unitary class, DMPK equation
}

\begin{document}
\sloppy
\maketitle

\section{Introduction}

The discovery of a perfectly conducting channel in disordered
wires~\cite{ando,takane0,takane1,takane2,takane3,sakai1,sakai2,sakai3,
barnes,hirose,wakabayashi,takane4,takane5,takane6} provides a counterexample
to the conjecture that an ordinary quasi-one-dimensional quantum system
with disorder exhibits Anderson localization
(i.e., conductance decays exponentially with increasing system length $L$
and eventually vanishes in the limit of $L \to \infty$).
We have shown that perfectly conducting channels can be stabilized in two
standard universality classes.
One is the symplectic universality class with an odd number of conducting
channels,~\cite{ando,takane0,takane1,
takane2,takane3,sakai1,sakai2,sakai3}
and the other is the unitary universality class with the imbalance between
the numbers of conducting channels in two propagating
directions.~\cite{barnes,hirose,wakabayashi,takane4,takane5,takane6}
The symplectic class consists of systems having time-reversal symmetry
without spin-rotation invariance, while the unitary class is characterized
by the absence of time-reversal symmetry.~\cite{beenakker1}

Much attention has recently been paid to electron transport
in the above-mentioned two universality classes
which do not exhibit Anderson localization.
For the symplectic class, one perfectly conducting channel is stabilized
in the odd-channel case, while such a special channel does not exist
in the ordinary even-channel case.
We have studied in details how the dimensionless
conductance $g$ behaves as a function of $L$.~\cite{sakai2,sakai3}
It is shown that the behavior of $g$ in the odd-channel case
is very different from that in the even-channel case
when $L$ is much longer than the conductance decay length $\xi$.
The dimensionless conductance in the even-channel case decays as $g \to 0$
with increasing $L$, while $g \to 1$ in the odd-channel case
due to the presence of one perfectly conducting channel.
Furthermore, the decay of $g$ with increasing $L$ is much faster
in the odd-channel case than in the even-channel case.
However, such a notable even-odd difference does not appear
in the diffusive regime of $L \ll \xi$.~\cite{sakai1,sakai2}
For the unitary class, the number of perfectly conducting channels depends on
the channel-number imbalance between two propagating directions.
If $m$ perfectly conducting channels are present, the dimensionless
conductance behaves as $g \to m$ with increasing $L$.
The present authors have shown that the notable $m$-dependence of $g$
appears in the long-$L$ regime, by using a scaling approach~\cite{takane4}
and a supersymmetry approach.~\cite{takane5,takane6}
However, the behavior of $g$ in the diffusive regime has not been clarified.
It is interesting to study whether the perfectly conducting channels affect
the behavior of $g$ in the diffusive regime.

In this paper, we focus on the unitary universality class
with the channel-number imbalance
and consider the conductance in the diffusive regime.
We here present the basic framework to describe the electron transport in
systems with the channel-number imbalance
on the basis of the scaling approach.~\cite{takane4}
Let us consider disordered wires with $N$ right-moving channels
and $N + m$ left-moving channels.
In this case, $m$ left-moving channels become perfectly conducting,
and the dimensionless conductances $g$ and $g'$ for the left-moving and
right-moving channels, respectively, differ from each other.
Each dimensionless conductance is determined by a corresponding set of
transmission eigenvalues.
If the set of the transmission eigenvalues for the right-moving channels is
$\{T_{1}, T_{2}, \dots, T_{N}\}$, that for the left-moving channels is
expressed as $\{T_{1}, T_{2}, \dots, T_{N}, 1, \dots, 1 \}$,
where we have identified the $N+1$ to $N+m$th channels as
the perfectly conducting ones.
The dimensionless conductance $g$ is given by
$g = \sum_{a=1}^{N+m}T_{a} = m + \sum_{a=1}^{N}T_{a}$,
while $g' = \sum_{a=1}^{N}T_{a}$.
We observe that $g = g' + m$.
It should be noted that the mean free paths $l$ and $l'$ for the left-moving
and right-moving channels, respectively, are not equal
due to the channel-number imbalance.
Indeed, they satisfy $l'= (N/(N+m))l$.
The statistical behavior of $g$, as well as $g'$, is described by
the probability distribution for $\{T_{1}, T_{2}, \dots, T_{N}\}$.
We define $\lambda_{a} \equiv (1-T_{a})/T_{a}$
and introduce the probability distribution $P(\{\lambda_{a}\};s)$
for $\{\lambda_{1}, \lambda_{2}, \dots, \lambda_{N}\}$,
where $s$ is the normalized system length defined by $s \equiv L/l$.
The Fokker-Planck equation for $P(\{\lambda_{a}\};s)$, which is called
the Dorokhov-Mello-Pereyra-Kumar (DMPK) equation,~\cite{dorokhov,mello1}
is expressed as~\cite{takane4}
\begin{align}
      \label{eq:dmpk-lambda}
 & \frac{\partial P(\{\lambda_{a}\};s)}{\partial s}
        \nonumber \\
 & =
   \frac{1}{N}\sum_{a=1}^{N} \frac{\partial}{\partial \lambda_{a}}
   \left( \lambda_{a}(1+\lambda_{a}) J
          \frac{\partial}{\partial \lambda_{a}}
            \left( \frac{P(\{\lambda_{a}\};s)}{J} \right)
   \right) 
\end{align}
with
\begin{equation}
      \label{eq:jacob}
  J = \prod_{c=1}^{N} \lambda_{c}^{m}
      \times
      \prod_{b=1}^{N-1}\prod_{a=b+1}^{N}|\lambda_{a}-\lambda_{b}|^{2} .
\end{equation}
The factor $\prod_{c=1}^{N} \lambda_{c}^{m}$ in $J$ represents the repulsion
arising from the $m$-fold degenerate perfectly conducting eigenvalue.
This eigenvalue repulsion suppresses the non-perfectly conducting eigenvalues
$\{T_{1}, T_{2}, \dots, T_{N}\}$.
It should be emphasized that all the influences of the perfectly conducting
channels are described by this factor.

The purpose of this paper is to study how the perfectly conducting channels
affect the dimensionless conductance $g$ in the diffusive regime.
Particularly, we focus on the variance,
${\rm var}\{g\} \equiv \langle g^{2} \rangle - \langle g \rangle^{2}$,
which characterizes the magnitude of conductance fluctuations.
We know that in the ordinary case of $m = 0$, the variance takes
the universal value $1/15$
irrespective of the normalized system length $s \equiv L/l$.
This is called universal conductance fluctuations.
Does this universality hold even
in the presence of the perfectly conducting channels?
To answer this question, we obtain ${\rm var}\{g\}$
by using two approaches based on the DMPK equation.
First, we analytically calculate the variance as a function of $s$
by using an $N^{-1}$ expansion approach.
This is applicable to the case in which $N \gg \max \{m,1\}$.
We show that the variance is given by ${\rm var}\{g\} = 1/15$
irrespective of $m$ in the diffusive regime of $N \gg s \gg 1$.
This indicates that in the large-$N$ limit, the universality of conductance
fluctuations hold even in the case of $m \neq 0$.
To study the case in which the ratio $m/N$ is not very small,
we numerically calculate ${\rm var}\{g\}$
by using a classical Monte Carlo approach based on
an approximate probability distribution for transmission eigenvalues.
This is applicable to the case of an arbitrary $m$ as long as $s/4N \ll 1$.
We treat the cases of $N = 6$ and $20$ with $m = 0,1,2,3$.
We show that in the case of $m = 0$, the variance approximately takes
a constant value nearly equal to $1/15$ for $N \gtrsim s \gtrsim 1$.
However, deviation from this universal behavior arises when $m \neq 0$.
We show that for $m \neq 0$, the variance does not take a constant value,
but decreases with increasing $s$,
and the corresponding deviation becomes more pronounced with increasing $m$.
We also show that the deviation in the case of $N = 6$ is more noteworthy
than that in the case of $N = 20$.
These results indicate that the universality of conductance fluctuations
breaks down for $m \neq 0$ unless $N$ is very large.

In the next section, we analytically obtain ${\rm var}\{g\}$
by using the $N^{-1}$ expansion approach.
In \S 3, we numerically calculate ${\rm var}\{g\}$
by using the classical Monte Carlo approach.
The Monte Carlo results are compared with those obtained in \S 2.
Section 4 is devoted to summary.

\section{$N^{-1}$ Expansion Approach}
We consider
${\rm var}\{g\} \equiv \langle g^{2} \rangle - \langle g \rangle^{2}$
in the diffusive regime of $N \gg s \gg 1$.
The ensemble average of an arbitrary function $F(\{\lambda_{a}\})$
is defined by
\begin{align}
      \label{eq:F-ave}
  \langle F \rangle
      = \int_{0}^{\infty} {\rm d}\lambda_{1}\cdots {\rm d}\lambda_{N}
        F(\{\lambda_{a}\}) P(\{\lambda_{a}\};s) .
\end{align}
Let $\Gamma$ be
\begin{align}
      \Gamma & \equiv \sum_{a=1}^{N} T_{a}
                    = \sum_{a=1}^{N}\frac{1}{1+\lambda_{a}} .
\end{align}
We can express ${\rm var}\{g\}$ in terms of
$\langle \Gamma^{p} \rangle$ with $p = 1$ and $2$.
To obtain $\langle \Gamma^{p} \rangle$, we derive the evolution equation
for it on the basis of the DMPK equation.
Combining the DMPK equation with eq.~(\ref{eq:F-ave}), we obtain
\begin{align}
     \label{eq:F-evolution}
 & N \frac{\partial \langle F \rangle}{\partial s}
              \nonumber \\
 & \hspace{5mm} = \left\langle
          \sum_{a=1}^{N}\frac{1}{J}\frac{\partial}{\partial \lambda_{a}}
          \left\{ \lambda_{a}(1+\lambda_{a})J
                  \frac{\partial F}{\partial \lambda_{a}}
          \right\}
     \right\rangle
              \nonumber \\
 & \hspace{5mm} = \Biggl\langle
        \sum_{a=1}^{N}
        \biggl\{ \lambda_{a}(1+\lambda_{a})
                 \frac{\partial^{2}F}{\partial \lambda_{a}^{2}}
               + (1+2\lambda_{a})\frac{\partial F}{\partial \lambda_{a}}
             \nonumber \\
 & \hspace{10mm}
               + \lambda_{a}(1+\lambda_{a})
                 \Biggl( \sum_{\scriptstyle b=1 \atop \scriptstyle (b \neq a)}
                                              ^{N}
                         \frac{2}{\lambda_{a}-\lambda_{b}}
                       + \frac{m}{\lambda_{a}}
                 \Biggr)\frac{\partial F}{\partial \lambda_{a}}
        \biggr\}
     \Biggr\rangle .
\end{align}
Replacing $F$ by $\Gamma^{p}$, we obtain
\begin{align}
     \label{eq:gamma_p}
  N \frac{\partial \langle \Gamma^{p} \rangle}{\partial s}
  & = - mp \langle \Gamma^{p} \rangle
           \nonumber \\
  & \hspace{-5mm}
       + p(p-1) \langle \Gamma^{p-2}\left(\Gamma_{2}-\Gamma_{3} \right) \rangle
      - p \langle \Gamma^{p+1} \rangle ,
\end{align}
where
\begin{align}
  \Gamma_{q} = \sum_{a=1}^{N}\frac{1}{\left(1+\lambda_{a}\right)^{q}} .
\end{align}

Note that eq.~(\ref{eq:gamma_p}) is not closed since its right-hand side
contains $\langle \Gamma^{p-2}\Gamma_{2}\rangle$
and $\langle \Gamma^{p-2}\Gamma_{3}\rangle$.
Even if eq.~(\ref{eq:gamma_p}) is combined with differential equations for
$\langle\Gamma^{p}\Gamma_{2}\rangle$ and $\langle\Gamma^{p}\Gamma_{3}\rangle$,
they do not form a closed set of equations and we cannot obtain
$\langle \Gamma^{p} \rangle$ in a simple manner.
To overcome this difficulty, we adapt the $N^{-1}$ expansion approach
presented by Mello and Stone.~\cite{mello2} 
This approach is adaptable if $N \gg \max \{m,1\}$.
We expand $\langle \Gamma^{p} \rangle$ in a power series of $N^{-1}$,
and obtain it up to order of $N^{-2+p}$.
To do so, we must supplement eq.~(\ref{eq:gamma_p}) by the following equations,
\begin{align}
         \label{eq:gamma_p2}
  N \frac{\partial \langle \Gamma^{p}\Gamma_{2} \rangle}{\partial s}
  & = - m(p+2) \langle \Gamma^{p}\Gamma_{2} \rangle
      + 4p \langle \Gamma^{p-1}\left(\Gamma_{3}-\Gamma_{4} \right) \rangle
           \nonumber \\
  & \hspace{5mm}
      + p(p-1) \langle \Gamma^{p-2}
               \left(\Gamma_{2}^{2}-\Gamma_{2}\Gamma_{3} \right) \rangle
             \nonumber \\
  & \hspace{5mm}
      + 2 \langle \Gamma^{p+2} \rangle
      - (p+4) \langle \Gamma^{p+1}\Gamma_{2} \rangle ,
                       \\
         \label{eq:gamma_p3}
  N \frac{\partial \langle \Gamma^{p}\Gamma_{3} \rangle}{\partial s}
  & = - m(p+3) \langle \Gamma^{p}\Gamma_{3} \rangle
      + 6p \langle \Gamma^{p-1}\left(\Gamma_{4}-\Gamma_{5} \right) \rangle
           \nonumber \\
  & \hspace{5mm}
      + p(p-1) \langle \Gamma^{p-2}
               \left(\Gamma_{2}\Gamma_{3}-\Gamma_{3}^{2} \right) \rangle
             \nonumber \\
  & \hspace{5mm}
      + 6 \langle \Gamma^{p+1}\Gamma_{2} \rangle
      - (p+6) \langle \Gamma^{p+1}\Gamma_{3} \rangle
           \nonumber \\
  & \hspace{5mm}
      - 3 \langle \Gamma^{p}\Gamma_{2}^{2} \rangle ,
                       \\
         \label{eq:gamma_p2-2}
  N \frac{\partial \langle \Gamma^{p}\Gamma_{2}^{2} \rangle}{\partial s}
  & = - m(p+4) \langle \Gamma^{p}\Gamma_{2}^{2} \rangle
      + 8 \langle \Gamma^{p}\left(\Gamma_{4}-\Gamma_{5} \right) \rangle
           \nonumber \\
  & \hspace{5mm}
      + 8p \langle \Gamma^{p-1}
               \left(\Gamma_{2}\Gamma_{3}-\Gamma_{2}\Gamma_{4} \right) \rangle
             \nonumber \\
  & \hspace{5mm}
      + p(p-1) \langle \Gamma^{p-2}
               \left(\Gamma_{2}^{3}-\Gamma_{2}^{2}\Gamma_{3} \right) \rangle
           \nonumber \\
  & \hspace{5mm}
      + 4 \langle \Gamma^{p+2}\Gamma_{2} \rangle
      - (p+8) \langle \Gamma^{p+1}\Gamma_{2}^{2} \rangle .
\end{align}
These equations can be derived from eq.~(\ref{eq:F-evolution}).
We employ the following expansions,
\begin{align}
        \label{eq:gp}
 & \langle \Gamma^{p} \rangle
   = N^{p}f_{p,0}(s) + N^{p-1}f_{p,1}(s) + N^{p-2}f_{p,2}(s) + \cdots ,
             \\
        \label{eq:gp2}
 & \langle \Gamma^{p}\Gamma_{2} \rangle
   = N^{p+1}j_{p+1,0}(s) + \cdots ,
             \\
        \label{eq:gp3}
 & \langle \Gamma^{p}\Gamma_{3} \rangle
   = N^{p+1}k_{p+1,0}(s) + \cdots ,
             \\
        \label{eq:gp2-2}
 & \langle \Gamma^{p}\Gamma_{2}^{2} \rangle
   = N^{p+2}l_{p+2,0}(s) + \cdots
\end{align}
with
\begin{align}
        \label{eq:ini-conditions}
  f_{p,n}(0) = j_{p,n}(0) = k_{p,n}(s) = l_{p,n}(0) = \delta_{n,0} .
\end{align}
Substituting these expansions into
eqs.~(\ref{eq:gamma_p}) and (\ref{eq:gamma_p2})-(\ref{eq:gamma_p2-2}),
and equating the coefficients of the various powers of $N$,
we obtain a set of closed differential equations for $f_{p,0}(s)$,
$f_{p,1}(s)$, $f_{p,2}(s)$, $j_{p,0}(s)$, $k_{p,0}(s)$ and $l_{p,0}(s)$.
We solve the resulting equations under the initial conditions
given in eq.~(\ref{eq:ini-conditions}).
These procedures are outlined in Appendix.
We finally obtain
\begin{align}
     \label{eq:gamma_p_result}
  \langle \Gamma^{p} \rangle
  & = \frac{N^{p}}{(1+s)^{p}} - \frac{mpN^{p-1}}{2(1+s)^{p+1}}(s^{2}+2s)
           \nonumber \\
  & \hspace{0mm}
      + \frac{m^{2}pN^{p-2}}{24(1+s)^{p+2}}
             \left((3p-1)s^{4}+(12p-4)s^{3}+12ps^{2}\right)
            \nonumber \\
  & \hspace{0mm} 
      + \frac{pN^{p-2}}{90(1+s)^{p+4}}
             \bigl((3p-5)s^{6}+(18p-30)s^{5}
           \nonumber \\
  & \hspace{5mm}
       +(45p-75)s^{4}+(60p-90)s^{3}+(45p-45)s^{2}\bigr) .
\end{align}
The variance of $\Gamma$, which is equal to ${\rm var}\{g\}$,
can be derived from this expression.
Up to order of $N^{0}$, we obtain
\begin{align}
  {\rm var} \{ \Gamma \}
     = \frac{1}{15} \frac{s^{6}+6s^{5}+15s^{4}+15s^{3}}{(1+s)^{6}} .
\end{align}
This results in
\begin{align}
 {\rm var}\{ g \} = \frac{1}{15}
\end{align}
irrespective of $m$ in the diffusive regime of $N \gg s \gg 1$.
This indicates that the perfectly conducting channels do not affect
${\rm var}\{g\}$ in the large-$N$ limit.

\section{Monte Carlo Approach}
To obtain ${\rm var}\{g\}$ without the restriction of $N \gg m$,
we employ a Monte Carlo approach based on an analytic expression
of the probability distribution for transmission eigenvalues.
Let us introduce a set of variables $x_{a}$, related to $\lambda_{a}$ by
$\lambda_{a} = \sinh^{2} x_{a}$.
We analytically obtain a simple approximate expression of the probability
distribution $P(\{x_{a}\};s)$ from the exact solution of the DMPK equation.
The DMPK equation has been solved exactly
for the ordinary case of $m = 0$.~\cite{beenakker2}
The exact solution for an arbitrary $m$ has been obtained
in ref.~\citen{akuzawa}.
In the notation which is convenient for our purpose,
the exact probability distribution is given by
\begin{align}
      \label{eq:prob_dis}
   P(\{x_{a}\};s)
 & = {\rm const.}
     \prod_{a,b=1 (a>b)}^{N} \left( \sinh^{2}x_{a} - \sinh^{2}x_{b} \right)
           \nonumber \\
  & \hspace{5mm}
     \times \prod_{a=1}^{N}\sinh2x_{a}\sinh^{2m}x_{a}
             \nonumber \\
 &  \hspace{5mm} \times
    {\rm det}\left\{ I_{b}(-\sinh^{2}x_{a})
             \right\}_{a,b = 1,2,\dots,N} ,
\end{align}
where ${\rm det}\{A_{ab} \}_{a,b = 1,2,\dots,N}$ denotes the determinant of
the $N \times N$ matrix $\hat{A}$ and
\begin{align}
     \label{eq:int_ab}
  I_{b}(-\sinh^{2}x_{a})
 & = \int_{0}^{\infty}{\rm d}k k^{2(b-1)}c_{m}^{2}(k)
                     {\rm e}^{-\frac{k^{2}}{4N}s}
           \nonumber \\
 & \hspace{5mm} \times F_{m}(k,-\sinh^{2}x_{a})
\end{align}
with
\begin{align}
 &  c_{m}^{2}(k)
     = \frac{1}{4\pi}
       \frac{\left|\Gamma\left(\frac{m+1+{\rm i}k}{2}\right)\right|^{4}}
            {\Gamma^{2}(m+1)|\Gamma({\rm i}k)|^{2}} ,
          \\
 & F_{m}(k,-\sinh^{2}x)
            \nonumber \\
 & \hspace{5mm} 
    = F \left(\frac{m+1-{\rm i}k}{2}, \frac{m+1+{\rm i}k}{2}, m+1; -\sinh^{2}x
       \right) .
\end{align}
When $s/4N \ll 1$, the dominant contribution to the integration over $k$
comes from the large-$k$ region of $k \gtrsim \sqrt{4N/s} \gg 1$.
In this region, we can approximate that
\begin{align}
       \label{eq:c_mk_approx}
 & c_{m}^{2}(k)
     = \frac{1}{2^{2m+1}(m!)^{2}} k^{2m+1} ,
          \\
       \label{eq:def-F}
 & F_{m}(k,-\sinh^{2}x)
            \nonumber \\
 & \hspace{5mm} 
    = F \left(\frac{1-{\rm i}k}{2}, \frac{1+{\rm i}k}{2}, m+1; -\sinh^{2}x
       \right) .
\end{align}
We express the asymptotic form of $F_{m}(k,-\sinh^{2}x)$
in terms of a Bessel function.
We start with the following expression
\begin{align}
     \label{eq:F-to-P}
 & F \left(\frac{1-{\rm i}k}{2}, \frac{1+{\rm i}k}{2}, m+1; -\sinh^{2}x \right)
           \nonumber \\
 & \hspace{15mm}
    = m! \left(\frac{\cosh x}{\sinh x}\right)^{m}
      P_{\frac{{\rm i}k-1}{2}}^{-m}(\cosh 2x) .
\end{align}
We replace the Legendre function by its asymptotic form
\begin{align}
     \label{eq:P-asympto}
 & P_{\frac{{\rm i}k-1}{2}}^{-m}(\cosh 2x)
    \sim \left(\frac{2}{k}\right)^{m+\frac{1}{2}}
         \left(\frac{2}{\pi \sinh 2x}\right)^{\frac{1}{2}}
            \nonumber \\
 & \hspace{30mm}
      \times \cos\left(kx-\frac{m\pi}{2}-\frac{\pi}{4}\right)
\end{align}
for large $k$.
This expression can be derived from eq.~(8.723) of ref.~\citen{gradshteyn}.
Combining eqs.~(\ref{eq:def-F})-(\ref{eq:P-asympto})
and the asymptotic form of the Bessel function
$J_{m}(y) \sim (2/\pi y)^{1/2}\cos(y-m\pi/2-\pi/4)$ for large $y$,
we arrive at
\begin{align}
     \label{eq:F_approx}
 F_{m}(k,-\sinh^{2}x)
 & \sim m! \left(\frac{2}{k}\right)^{m}
        \left(\frac{\cosh x}{\sinh x}\right)^{m}
        \left(\frac{2x}{\sinh 2x}\right)^{\frac{1}{2}}
            \nonumber \\
 & \hspace{15mm} \times J_{m}(kx) .
\end{align}
Substituting eqs.~(\ref{eq:c_mk_approx}) and (\ref{eq:F_approx}) into
eq.~(\ref{eq:int_ab}) and carrying out the $k$-integration, we obtain
\begin{align}
      \label{eq:int_ab_result}
  I_{b}(-\sinh^{2}x_{a})
 & = \frac{(b-1)!}{2^{2m+2}m! \left(\frac{s}{4N} \right)^{b+m}}
    \left(\frac{x_{a}\cosh x_{a}}{\sinh x_{a}}\right)^{m}
            \nonumber \\
 & \hspace{5mm} 
    \times \left(\frac{2x_{a}}{\sinh 2x_{a}}\right)^{\frac{1}{2}}
    {\rm e}^{-\frac{Nx_{a}^{2}}{s}}
    L_{b-1}^{m}\left(\frac{Nx_{a}^{2}}{s}\right) ,
\end{align}
where $L_{b-1}^{m}$ is the Laguerre polynomial.
Substituting eq.~(\ref{eq:int_ab_result}) into eq.~(\ref{eq:prob_dis}) and
using the following relation
\begin{align}
 & {\rm det}\left\{ L_{b-1}^{m}\left(\frac{Nx_{a}^{2}}{s}\right)
           \right\}_{a,b = 1,2,\dots,N}
            \nonumber \\
 & \hspace{15mm} 
  = {\rm const.} \prod_{a,b=1(a > b)}^{N} \left(x_{a}^{2}-x_{b}^{2}\right) ,
\end{align}
we obtain
\begin{align}
      \label{eq:prob_dis_result}
 &  P(\{x_{a}\};s)
            \nonumber \\
 & \hspace{5mm} 
   = {\rm const.} \prod_{a,b=1(a > b)}^{N}
     \left\{ \left(\sinh^{2}x_{a} - \sinh^{2}x_{b}\right)
             \left(x_{a}^{2} - x_{b}^{2}\right)
     \right\}
            \nonumber \\
 & \hspace{15mm} \times
     \prod_{a=1}^{N}
     \left\{ \left(x_{a}\sinh 2x_{a}\right)^{m+\frac{1}{2}}
     {\rm e}^{-\frac{Nx_{a}^{2}}{s}}
     \right\} .
\end{align}
For the case of $m = 0$, the identical probability distribution
has been obtained in ref.~\citen{beenakker2}.
It is convenient to rewrite eq.~(\ref{eq:prob_dis_result}) as
\begin{align}
      \label{eq:prob_dis_mod}
   P(\{x_{a}\};s)
 = {\rm const.} \, {\rm e}^{ - H(\{x_{a}\})} ,
\end{align}
where
\begin{align}
  H(\{x_{a}\})
   & = \sum_{a=1}^{N}
       \left(   \gamma x_{a}^{2}
              - \left(m+\frac{1}{2}\right)
                \ln \left|x_{a}\sinh 2x_{a} \right|
       \right)
          \nonumber \\
   & \hspace{-10mm}
     - \sum_{a,b=1(a>b)}^{N}
       \Bigl(   \ln \left|\sinh^{2}x_{a}-\sinh^{2}x_{b}\right|
              + \ln \left|x_{a}^{2}-x_{b}^{2}\right|
       \Bigr)
\end{align}
with $\gamma \equiv N/s$.
From eq.~(\ref{eq:prob_dis_mod}), we find that the average of
$\Gamma^{p} = (\sum_{a=1}^{N}1/\cosh^{2}x_{a})^{p}$ is expressed as
\begin{align}
  \langle \Gamma^{p} \rangle
   = Z^{-1} \int_{0}^{\infty} {\rm d}x_{1}\cdots {\rm d}x_{N}
            \Gamma^{p} {\rm e}^{- H(\{x_{a}\})}
\end{align}
with
\begin{align}
  Z = \int_{0}^{\infty} {\rm d}x_{1}\cdots {\rm d}x_{N}
      {\rm e}^{- H(\{x_{a}\})} .
\end{align}

Strictly speaking, our approach is justified only when $\gamma^{-1}$ is
sufficiently small, since we have assumed $s/4N \ll 1$ in the derivation of
eq.~(\ref{eq:prob_dis_result}).
However, we can expect that it is qualitatively reliable even when
$\gamma^{-1}$ is not small.
To see this, let us consider the large-$\gamma^{-1}$ limit.
In this limit, the variables $\{x_{a}\}$ become much greater than unity
and are widely separated with each other,
so that we can assume $1 \ll x_{1} \ll x_{2} \ll \cdots \ll x_{N}$.
Under this assumption, $H(\{x_{a}\})$ can be approximated as
\begin{align}
          \label{eq:large-gamma-H}
  H(\{x_{a}\})
     = \sum_{a=1}^{N}
       \left( \gamma x_{a}^{2} - (2a - 1 + \eta m) x_{a} \right)
\end{align}
with $\eta = 2$.
Equation~(\ref{eq:large-gamma-H}) is identical to the correct expression
in the large-$\gamma^{-1}$ limit~\cite{takane4} if we substitute $\eta = 1$.
This implies that the probability distribution
given in eq.~(\ref{eq:prob_dis_mod}) is qualitatively reliable
even for not small $\gamma^{-1}$
although it overestimates the influence of perfectly conducting channels.

Note that we can interpret $H(\{x_{a}\})$ as the Hamiltonian function of
$N$ classical particles in one dimension.
This analogy allows us to adapt a Monte Carlo approach
to numerical calculations of $\langle \Gamma^{p} \rangle$.
Using a simple Metropolis algorithm,~\cite{froufe-perez,canali}
we compute ${\rm var}\{g\}$ for $N = 6$ and $20$
with $m = 0,1,2,3$ as a function of $\gamma^{-1} = s/N$.
The results are shown in Fig.~1,
where $\gamma^{-1}$ is restricted to $1 \ge \gamma^{-1} \ge 0$
and the average for each data point is taken over
$5 \times 10^{7}$ Monte Carlo steps.
\begin{figure}[hbtp]
\includegraphics[height=6cm]{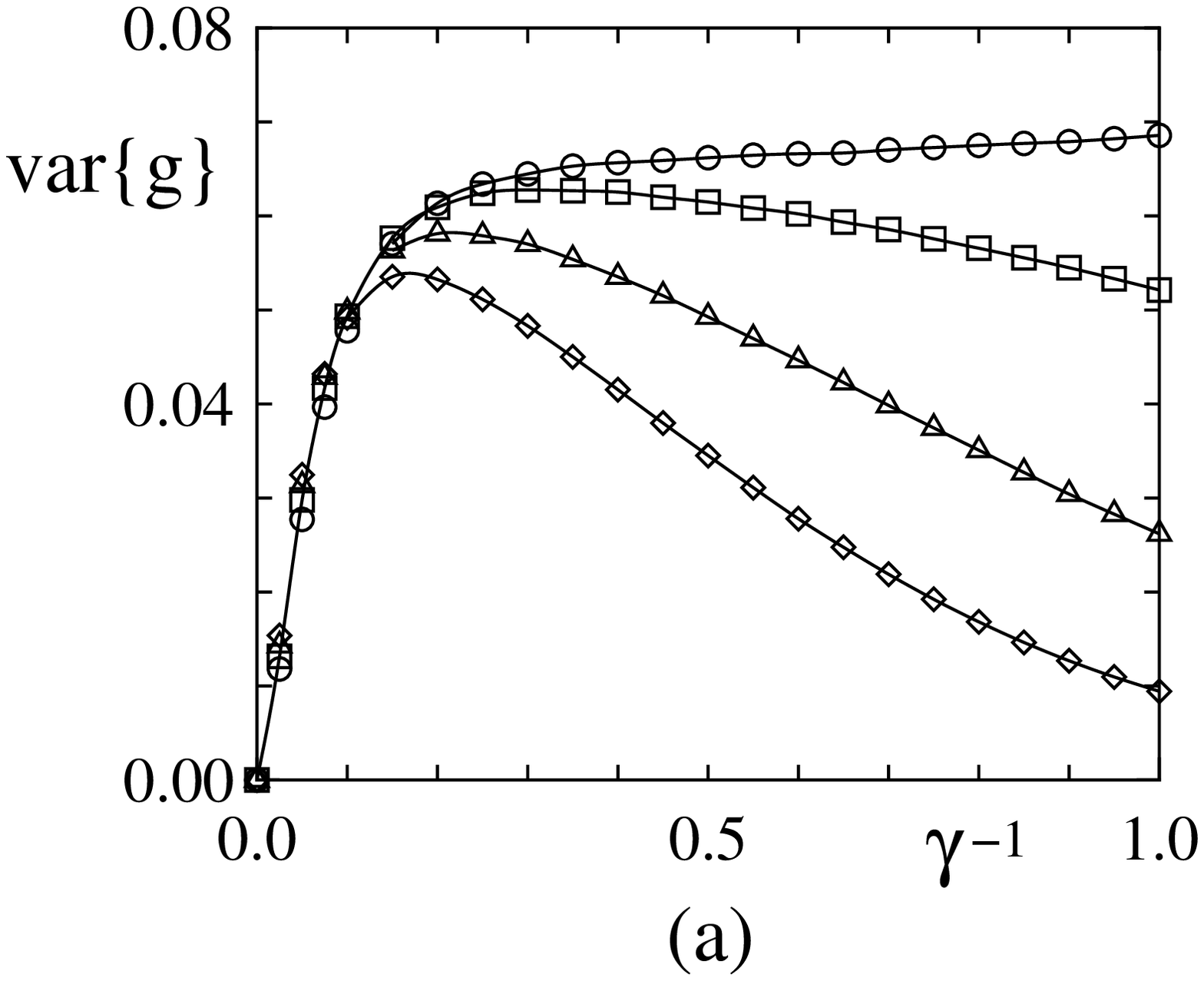}
\includegraphics[height=6cm]{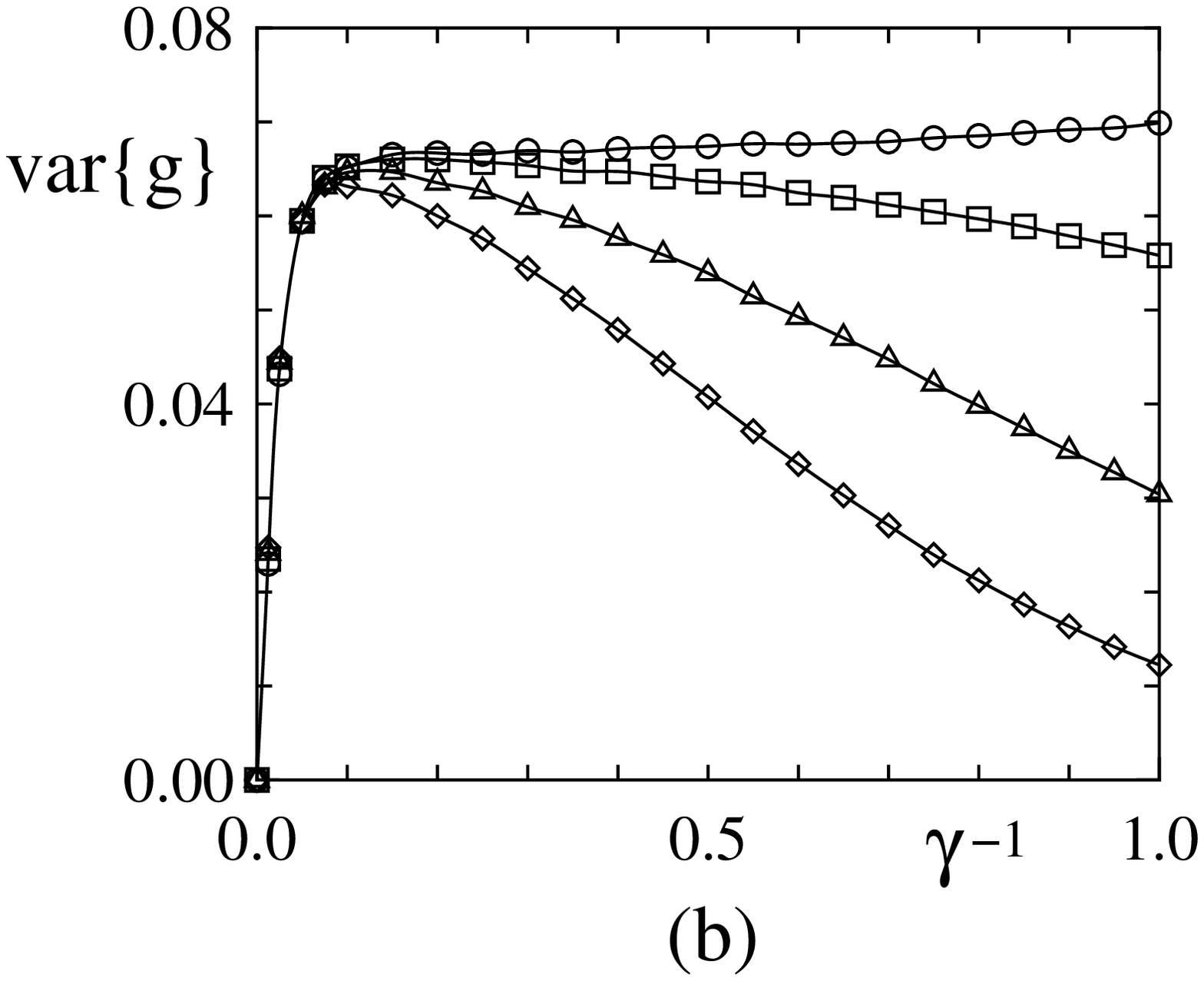}
\caption{The variance of $g$ for (a) $N = 6$ and (b) $N = 20$
as a function of $\gamma^{-1} \equiv s/N$.
Circles, squares, triangles and diamonds correspond to
$m = 0,1,2$ and $3$, respectively.
The solid lines are guide to eyes.
}
\end{figure}
Figure~1 shows that ${\rm var}\{g\}$ depends on $m$.
In the ordinary case of $m = 0$, we observe from this figure that
${\rm var}\{g\}$ increases with increasing $\gamma^{-1}$ for
$N^{-1} \gtrsim \gamma^{-1} \ge 0$ and approximately takes a constant
nearly equal to the universal value $1/15$ for $\gamma^{-1} \gtrsim N^{-1}$.
However, deviation from this universal behavior arises when $m \neq 0$.
For $m \neq 0$, we observe that ${\rm var}\{g\}$ does not take a constant
value, but decreases with increasing $\gamma^{-1}$
for $\gamma^{-1} \gtrsim N^{-1}$.
The decrease of ${\rm var}\{g\}$ becomes more pronounced with increasing $m$.
We also observe that the deviation in the case of $N = 6$ is more noteworthy
than that in the case of $N = 20$.
We conclude that the universality of conductance fluctuations breaks down
for $m \neq 0$ unless $N$ is very large.
This conclusion is not inconsistent with the result of \S 2,
because the $N^{-1}$ expansion approach is justified only
in the large-$N$ limit.

\section{Summary}

We have studied the variance of the dimensionless conductance $g$
in disordered wires with unitary symmetry in the diffusive regime.
We have focused on the case in which the number of left-moving conducting
channels is $N+m$, while that of right-moving ones is $N$.
In this case, $m$ left-moving channels become perfectly conducting.
First, we have analytically obtained the variance
as a function of $s = L/l$ by using an $N^{-1}$ expansion approach.
We have shown that the variance is given by ${\rm var}\{g\} = 1/15$
irrespective of $m$ when $N \gg \max \{m,1\}$.
This indicates that we cannot detect influences of the perfectly conducting
channels in the large-$N$ limit.
Second, we have numerically calculated ${\rm var}\{g\}$ by using
a classical Monte Carlo approach
to study the case in which the ratio $m/N$ is not very small.
We have shown that ${\rm var}\{g\}$ approximately takes a constant value
nearly equal to $1/15$ only in the ordinary case of $m = 0$, but
deviation from it appears when $m \ge 1$.
We have also shown that the deviation becomes more pronounced
with increasing $m$, particularly for small $N$.
We conclude that the universality of conductance fluctuations breaks down
for $m \neq 0$ unless $N$ is very large.

\section*{Acknowledgment}
K. W. acknowledges the financial support by a Grant-in-Aid for 
Young Scientists (B) (No. 19710082) from the Ministry of Education,
Culture, Sports, Science and Technology, also by a Grand-in-Aid for
Scientific Research (B) from the Japan Society for the Promotion of Science
(No. 19310094).

\appendix
\section{Derivation of $\langle\Gamma^{p}\rangle$}
We substitute eqs.(\ref{eq:gp})-(\ref{eq:gp3}) into eq.~(\ref{eq:gamma_p}) and
equate the coefficient of $N^{p+q}$ ($q=1,0,-1$) in the left-hand side
with that in the right-hand side.
We obtain
\begin{align}
      \label{eq:fp0}
 f'_{p,0}(s) & = - pf_{p+1,0}(s) ,
         \\
      \label{eq:fp1}
 f'_{p,1}(s) & = - mpf_{p,0}(s) - pf_{p+1,1}(s) ,
         \\
      \label{eq:fp2}
 f'_{p,2}(s) & = - mpf_{p,1}(s) + p(p-1)\left(j_{p-1,0}(s)-k_{p-1,0}(s)\right)
            \nonumber \\
             & \hspace{10mm} 
                 - pf_{p+1,2}(s) .
\end{align}
In a manner similar to this, we substitute eqs.~(\ref{eq:gp})-(\ref{eq:gp2-2})
into eqs.~(\ref{eq:gamma_p2})-(\ref{eq:gamma_p2-2}) and obtain
\begin{align}
      \label{eq:jp0}
 j'_{p,0}(s) & = 2 f_{p+1,0}(s) - (p+3) j_{p+1,0}(s) ,
         \\
      \label{eq:kp0}
 k'_{p,0}(s) & = 6 j_{p+1,0}(s) - (p+5) k_{p+1,0}(s) - 3 l_{p+1,0}(s),
         \\
      \label{eq:lp0}
 l'_{p,0}(s) & = 4 j_{p+1,0}(s) - (p+6) l_{p+1,0}(s) .
\end{align}
We obtain $\langle\Gamma^{p}\rangle$ by solving
eq.~(\ref{eq:fp0})-(\ref{eq:lp0}).

From eq.~(\ref{eq:fp0}), we obtain
\begin{align}
      \label{eq:fp0-result}
   f_{p,0}(s) = \frac{1}{(1+s)^{p}} .
\end{align}
Substituting this into eq.~(\ref{eq:fp1}), we obtain
\begin{align}
       \label{eq:fp1-mod1}
   f'_{p,1}(s) + pf_{p+1,1}(s) = - \frac{mp}{(1+s)^{p}} .
\end{align}
If we assume that $f_{p,1}(s) = p\chi_{p}(s)$ with
$\chi_{p+1}(s) = (1+s)^{-1}\chi_{p}(s)$,
the above equation is reduced to
\begin{align}
  \chi'_{p}(s) + \frac{p+1}{1+s}\chi_{p}(s) = - \frac{m}{(1+s)^{p}} .
\end{align}
Solving this equation with $f_{p,1}(0) = p\chi_{p}(0) = 0$,
we obtain $\chi_{p}(s) = - (m/2)(s^{2}+2s)/(1+s)^{p+1}$.
This results in
\begin{align}
       \label{eq:fp1-result}
   f_{p,1}(s) = - \frac{mp}{2}\frac{s^{2}+2s}{(1+s)^{p+1}} .
\end{align}

We need $j_{p-1,0}(s)$ and $k_{p-1,0}(s)$ to obtain $f_{p,2}(s)$.
Substituting eq.~(\ref{eq:fp0-result}) into eq.~(\ref{eq:jp0}),
we obtain
\begin{align}
       \label{eq:jp0-mod1}
   j'_{p,0}(s) + (p+3)j_{p+1,0}(s) = \frac{2}{(1+s)^{p+1}} .
\end{align}
If we assume that $j_{p,0}(s)$ satisfies $j_{p+1,0}(s) = (1+s)^{-1}j_{p,0}(s)$,
eq.~(\ref{eq:jp0-mod1}) is reduced to
\begin{align}
       \label{eq:jp0-mod2}
   j'_{p,0}(s) + \frac{p+3}{1+s} j_{p,0}(s) = \frac{2}{(1+s)^{p+1}} .
\end{align}
The solution of eq.~(\ref{eq:jp0-mod2}) with $j_{p,0}(0) = 1$ is given by
\begin{align}
       \label{eq:jp0-result}
   j_{p,0}(s) = \frac{2s^{3}+6s^{2}+6s+3}{3(1+s)^{p+3}} .
\end{align}
We here treat $l_{p,0}(s)$ which is necessary to obtain $k_{p,0}(s)$.
Substituting eq.~(\ref{eq:jp0-result}) into eq.~(\ref{eq:lp0})
and assuming that $l_{p,0}(s)$ satisfies $l_{p+1,0}(s) = (1+s)^{-1}l_{p,0}(s)$,
we obtain
\begin{align}
       \label{eq:lp0-mod1}
 & l'_{p,0}(s) + \frac{p+6}{1+s} l_{p,0}(s)
            \nonumber \\
 & \hspace{10mm} 
     = \frac{4}{3(1+s)^{p+4}}\left( 2s^{3}+6s^{2}+6s+3 \right) .
\end{align}
The solution of eq.~(\ref{eq:lp0-mod1}) with $l_{p,0}(0) = 1$ is given by
\begin{align}
       \label{eq:lp0-result}
   l_{p,0}(s)
     & = \frac{1}{9(1+s)^{p+6}}
         \bigl( 4s^{6}+24s^{5}+60s^{4}+84s^{3}
            \nonumber \\
     & \hspace{25mm} 
            +72s^{2}+36s+9 \bigr) .
\end{align}
We substitute eqs.~(\ref{eq:jp0-result}) and (\ref{eq:lp0-result})
into eq.~(\ref{eq:kp0})
and assume that $k_{p,0}(s)$ satisfies $k_{p+1,0}(s) = (1+s)^{-1}k_{p,0}(s)$.
We then obtain the following differential equation
\begin{align}
       \label{eq:kp0-mod1}
 &  k'_{p,0}(s) + \frac{p+5}{1+s} k_{p,0}(s)
            \nonumber \\
 & \hspace{5mm} 
     = \frac{1}{3(1+s)^{p+7}}
       \bigl( 8s^{6}+48s^{5}+120s^{4}+162s^{3}
            \nonumber \\
 & \hspace{35mm} 
    +126s^{2}+54s+9 \bigr) .
\end{align}
The solution of eq.~(\ref{eq:kp0-mod1}) with $k_{p,0}(0) = 1$ is given by
\begin{align}
       \label{eq:kp0-result}
 & k_{p,0}(s)
      = \frac{1}{15(1+s)^{p+6}}
        \bigl( 8s^{6}+48s^{5}+120s^{4}+165s^{3}
             \nonumber \\
 & \hspace{37mm} 
           +135s^{2}+60s+15 \bigr) .
\end{align}

We now turn to the evaluation of $f_{p,2}(s)$.
Substituting eqs.~(\ref{eq:fp1-result}), (\ref{eq:jp0-result})
and (\ref{eq:kp0-result}) into eq.~(\ref{eq:fp2}), we obtain
\begin{align}
       \label{eq:fp2-mod1}
   f'_{p,2}(s) + pf_{p+1,2}(s)
  & = \frac{m^{2}p^{2}}{2(1+s)^{p+1}} \left( s^{2}+2s \right)
           \nonumber \\
  & \hspace{-20mm}
      + \frac{p(p-1)}{15(1+s)^{p+5}}
        \bigl( 2s^{6}+12s^{5}+30s^{4}+40s^{3}
            \nonumber \\
  & \hspace{15mm} 
        +30s^{2}+15s \bigr) .
\end{align}
If we decompose $f_{p,2}(s)$ as $f_{p,2}(s) = m^{2}u_{p}(s)+v_{p}(s)$,
we obtain the differential equations for $u_{p}(s)$ and $v_{p}(s)$ as
\begin{align}
     \label{eq:u}
  u'_{p}(s) + p u_{p+1}(s)
   & = \frac{p^{2}}{2(1+s)^{p+1}}\left( s^{2}+2s \right) ,
         \\
     \label{eq:v}
  v'_{p}(s) + p v_{p+1}(s)
   & = \frac{p(p-1)}{15(1+s)^{p+5}}
       \bigl( 2s^{6}+12s^{5}+30s^{4}
            \nonumber \\
   & \hspace{10mm} 
          +40s^{3}+30s^{2}+15s \bigr) .
\end{align}
We can solve eq.~(\ref{eq:u}) by assuming
$u_{p}(s) = p^{2}\mu_{p}(s) +p\nu_{p}(s)$ with
$\mu_{p+1}(s)=(1+s)^{-1}\mu_{p}(s)$ and $\nu_{p+1}(s)=(1+s)^{-1}\nu_{p}(s)$.
Equation~(\ref{eq:v}) can also be solved by assuming that
$v_{p}(s) = p^{2}\xi_{p}(s) +p\tau_{p}(s)$ with
$\xi_{p+1}(s)=(1+s)^{-1}\xi_{p}(s)$ and $\tau_{p+1}(s)=(1+s)^{-1}\tau_{p}(s)$.
Combining the resulting $u_{p}(s)$ and $v_{p}(s)$,
we obtain
\begin{align}
     \label{eq:fp2-result}
  f_{p,2}(s)
 & = \frac{m^{2}p}{24(1+s)^{p+2}}
     \left( (3p-1)s^{4}+(12p-4)s^{3}+12ps^{2} \right)
          \nonumber \\
 & \hspace{5mm}
   + \frac{p}{90(1+s)^{p+4}}
     \bigl( (3p-5)s^{6}+(18p-30)s^{5}
            \nonumber \\
 & \hspace{15mm} +(45p-75)s^{4}+(60p-90)s^{3}
            \nonumber \\
 & \hspace{15mm} +(45p-45)s^{2}  \bigr) .
\end{align}
Substituting eqs.~(\ref{eq:fp0-result}), (\ref{eq:fp1-result})
and (\ref{eq:fp2-result}) into eq.~(\ref{eq:gp}),
we finally obtain eq.~(\ref{eq:gamma_p_result}).

\end{document}